\documentclass[aps,prb,twocolumn,showpacs,showkeys]{revtex4}
\usepackage{natbib}
\usepackage[]{amsmath}
\usepackage{epsf}
\usepackage{psfrag}
\usepackage{mathrsfs}
\usepackage{graphicx}
\usepackage{dcolumn}
\usepackage{bm}

\begin{document}

\title{Quasiparticle generation efficiency in superconducting thin-films}
\author{T. Guruswamy, D. J. Goldie and S. Withington}
\affiliation {Detector and Optical Physics Group, Cavendish Laboratory, University of Cambridge,
J J Thomson Avenue, Cambridge, CB3 0HE, UK \\
}
\email{d.j.goldie@mrao.cam.ac.uk}

\date{\today}

\begin{abstract}
Thin-film superconductors with thickness $~\sim 30-500\,\,{\rm nm}$
 are used 
 as non-equilibrium quantum detectors for photons, phonons or more exotic particles.
One of the most basic questions in determining their
limiting sensitivity is the efficiency with which the quanta of interest couple to the detected quasiparticles.
As low temperature superconducting resonators,  thin-films are attractive candidates for producing quantum-sensitive
 arrayable  sensors
and the readout uses an additional  microwave probe.
%
  We have calculated the quasiparticle generation efficiency $\eta_s$ for low energy photons
in a representative, clean  thin-film
  superconductor (Al) operating well-below its superconducting transition temperature 
 as a function of film thickness,
within the framework of the coupled kinetic equations
described by Chang and Scalapino.[J. J. Chang and D. J. Scalapino, J. Low Temp. Phys. {\bf 31}, 1 (1978)].
 We have also included
 the effect of a lower frequency probe.
We show that  phonon loss from the thin-film reduces $\eta_s$ by as much as $40\%$
compared to earlier models that considered relatively thick films or infinite volumes.
We also show that the presence of the probe and signal {\it enhances} the generation efficiency slightly. 
We conclude that the ultimate limiting noise
equivalent power of this class of detector is determined by the thin-film geometry.
\end{abstract}
%

%
%
\maketitle

\section{Introduction}

 Superconductive  detectors have revolutionized experimental astrophysics. 
Many of these detectors exploit Cooper pair-breaking in a thin-film low transition temperature superconductor
operating at low reduced temperatures $T/T_c\simeq 0.1$
 ($T$ is the temperature and $T_c$ is the superconducting
transition temperature, $T_c\sim1\,\,{\rm   K}$).
 These detectors rely on  non-equilibrium effects but to our
knowledge no detailed microscopic description exists of 
the
efficiency with which the excess quasiparticles are created
in a thin-film superconductor of thickness $~\sim 30-500\,\,{\rm nm}$.
This problem is very relevant not just for kinetic inductance detectors  (KIDs),\cite{Jonas_review, Baselmans_review}
 but also for
superconducting tunnel junction detectors,\cite{Friedrich_review}
single photon counting nanowires,\cite{Goltsman_sspd}
and
 quantum capacitance detectors.\cite{Bueno_2010_qcd}
 All of these devices can be fabricated by photolithography usually on a relatively thick substrate such as Si or
sapphire that is held at $T$ and that functions as a heat bath.
KIDs  are thin-film superconducting resonators
that can be configured
as ultra-sensitive   detectors of signal photons across the electromagnetic spectrum. 
KIDs are typically readout with a microwave probe with photons of energy $h\nu_p \sim 0.05\Delta$,
 where $\nu_p$ is the probe frequency, $2\Delta$ is the low temperature  superconducting energy gap
and $h$ is Planck's constant.
In this instance understanding combined effects of the signal and the probe is  clearly important.
In Ref.~\onlinecite{Goldie_SuST_2013} we described a detailed microscopic calculation of the spectrum of the non-equilibrium quasiparticles and phonons in a
superconducting resonator operating at $T/T_c=0.1$ considering only a probe.
 Ref.~\onlinecite{de_Visser_Goldie_2013} compared  that model  with precise experimental
measurements of the temperature and power dependence of the behavior of ultra-sensitive Al resonators,
finding good agreement between model and measurement.  

In a superconductor each absorbed signal photon with energy $h\nu_s\ge2 \Delta$ breaks a pair ($\nu_s$ is the signal frequency).
Probably the most important consideration in calculating the detection sensitivity of
these photons in  {\it any} thin-film superconducting detector
 is $\eta_s$ the average fraction of the  photon energy
that creates low energy quasiparticles $E\sim\Delta$ where $E$ is the quasiparticle energy.
We distinguish the {\it primary} spectrum of quasiparticles generated by the signal from the  driven
{\it quasistatic}
population that is established as the primary spectrum relaxes temporally and energetically.
Absorbed photons create  a spectrum of excess primary quasiparticles  that relaxes
to energies $E\sim\Delta$ by emitting phonons on a  timescale $\tau_{\rm {cascade}} \sim 0.1\text {--} 10 \,\,{\rm ns}$,\cite{Kozorezov_2000, Kozorezov_2012}
determined by the quasiparticle-phonon scattering time
at $E=3\Delta$. $\tau_{\rm {cascade}}$   is much shorter than
the effective loss time from the film of the excess energy
contained in the  
quasistatic distribution by $2\Delta$-phonon loss.
This time  is determined by
the effective quasiparticle recombination time of the excess $\tau_r^{\rm {eff}}$
provided other relatively slow direct quasiparticle loss-mechanisms such as out-diffusion or tunneling can be ignored.
For
$T/T_c\sim 0.1$ and for low detected power, $\tau_r^{\rm {eff}}\sim\,\, {\rm ms}$  even in a thin Al film.\cite{de_Visser_Goldie_2013}
Since $\tau_r^{\rm {eff}}\gg \tau_{\rm {cascade}}$ the low energy  quasistatic
 population  determines the detector response.
During the energy relaxation pair-breaking by the emitted phonons
occurs provided $\Omega\ge2\Delta$ ($\Omega$ is the phonon energy) and this increases the quasistatic population near $\Delta$, although
phonon loss is also possible with characteristic time $\tau_{l}$.
 At low temperature and low phonon energies $\Omega\sim 2\Delta$, the pair-breaking time  $\tau_{pb}(\Omega)\sim\tau_0^\phi$ where
$\tau_0^\phi$ is the characteristic phonon lifetime.\cite{Kaplan}
We assume that $\tau_{l}$ is independent of $\Omega$ and is determined by the film thickness and the
coupling to the substrate.\cite{Kaplan_loss}
For thin-films $\tau_{l}$  is comparable with, or even less than
$\tau_{pb}$.
%
%
Phonon-loss means that energy is lost  from any finite thickness film
before the quasistatic population is established.

\begin{figure}[!t]
 \begin{center}
   \begin{tabular}{c}
     \psfrag{xaxis} [] [] { $E /\Delta  $ }
     \psfrag{yaxis} [] [] { $ \tau^{-1}\,\,{(\rm ns}^{-1})  $  }
     \psfrag{E1} [c] [c] [1] { $E = 3 \Delta $ }
     \psfrag{T1} [c] [c] [1] { $(\tau_{e\phi})^{-1}$ }
     \psfrag{T2} [c] [c] [1] { $(\tau_{ee})^{-1} $ }
     \psfrag{T3} [c] [c] [1] { $\left( \tau_{ee}\left(D\right)\right)^{-1}$}
   \includegraphics[height=8.5cm, angle=-90]{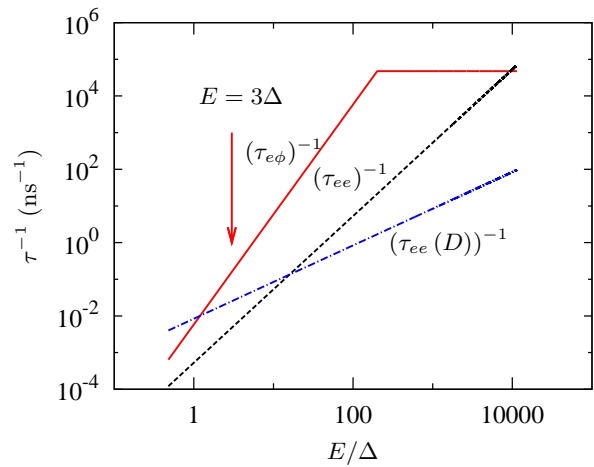}
   \end{tabular}
  \end{center}
  \caption[Fig1]
   { \label{fig:Rate_cf}
 (Color online) Energy dependence of the  relaxation rates in a clean, thin Al film in the normal state: (red) solid line electron-phonon scattering,
(black) dashed line clean-limit electron-electron scattering and (blue) dash-dot line the electron-electron scattering time including the effect of disorder. The calculations are for a
 $35\,\,{\rm nm}$ Al film with $\rho = 8 \times 10^{-8}\,\,{\rm \Omega m}$. $\Delta$ is the
 low temperature energy gap.
 }
   \end{figure}

Up to this point we have ignored the effect of the electron-electron interaction in the energy
down-conversion.
Figure~\ref{fig:Rate_cf} shows the energy dependence of the
{\it normal-state} scattering rates
 due to the electron-phonon
(e-$\phi$) interaction $\tau_{e\phi}^{-1}$,\cite{Kaplan}
the
 clean-limit electron-electron (e-e) rate $\tau_{ee}^{-1} $ which is also valid for
 disordered films at high energies,\cite{Kozorezov_2000, Quinn_and_Ferrell}
and the e-e rate including the effect of disorder $\left( \tau_{ee}\left(D\right)\right)^{-1}$,\cite{Gershenzon_1990, Altshuler_Aronov_1985}
where $D$ denotes the diffusion coefficient.
 The calculations  assume an Al thickness  $d = 35\,\,{\rm nm}$
  with
resistivity $\rho= 8\times 10^{-8}\,\,{\rm \Omega m}$, typical of a clean Al film on
 $\rm{Al_2 O_3}$,\cite{de_Visser_Goldie_2013}  that is representative of the thinnest
 films modeled here.
For $E>200\Delta$, $(\tau_{e\phi})^{-1}$ is cut-off at the
Debye energy $\Omega_D$. At the highest energies $E\sim  10^4\Delta$
e-e scattering becomes the
main energy relaxation mechanism. At lower energies ($E\sim 25 \Delta$) disorder increases the e-e rate
in this instance.  For
$E\sim 1.2 \Delta$ we see that the e-e and e-$\phi$ rates again become equal.
For $ 1.2 \Delta < E < 10^4\Delta$ the e-$\phi$ interaction is the principal relaxation mechanism.
A detailed description  of the  energy dependence   of the disorder-enhanced e-e rate
  in a superconductor at low $T/T_c$, including the effect of the
  energy gap,
seems to be lacking although for $E = \Delta$ the
e-e rate is further reduced compared to   e-$\phi$  becoming
negligible.\cite{Sergeev_Reizer_1996}
We note also that the energy scale of interest determining the relative importance of
low energy e-e compared to  e-$\phi$ scattering  in the relaxation
  is not  $\Delta$ but
rather $3\Delta$: below this energy pair-breaking is forbidden for both.
For the thicker Al films  discussed below $D$ is enhanced in clean films so that the effect of disorder is again reduced.
For these reasons we ignore e-e relaxation for all energy scales, temperatures and film parameters considered.
Extrapolation of our results to other low-$T_c$ superconductors  should thus be done with caution
particularly for
higher resistivity or very thin films.

A number of calculations exist of $\eta_s$ for high energy photons
$h\nu_s\gg 2\Delta$ and $\Omega_D$ that have considered infinite superconducting volumes 
finding $\eta_s\sim0.57\text {--\,} 0.6 $ for  Al,\cite{Kozorezov_2000}
 Nb\cite{Rando_1992} and Sn.\cite{Kurakado}
 Hijmering {\it et al.}\cite{Hijmering_2009} calculated quasiparticle creation efficiencies in thin-film Al-Ta bilayers taking account of the
modification of the quasiparticle density of states due to the proximity effect but  ignored loss
of pair-breaking phonons.
Zehnder\cite{Zehnder_model}
 calculated $\eta_s$    in a number of thin film superconductors with thickness $d=500\,\,{\rm nm}$ at
$T=0.5 \,\,{\rm K}$  including quasiparticle diffusion and phonon loss. $\eta_s$ was determined from the number of
quasiparticles remaining at time $t\simeq 10\,\,{\rm ns}$ when the initial energy down-conversion was considered complete
giving $\eta_s \sim 0.7$ for Al.

Here we consider the regime $90\le \nu_s\le 450\,\,{\rm GHz}$.
 To date no work has calculated $\eta_s$  for  these signal photon energies
 at low $T/T_c$, or the
technologically important range of film thicknesses considered here including $2\Delta$-phonon loss.
 This  frequency range is particularly relevant for mm- and sub-mm astronomy.
 We have also included a lower frequency probe.
We followed Chang and Scalapino\cite{Chang_and_Scalapino_ltp}
to solve the coupled kinetic equations describing the quasiparticle and phonon populations.
Our  approach
explicitly includes the contribution of  all phonon branches
 because it relies on  the measured Eliashberg function $\alpha^2 F(\Omega)$
in the calculation  of the characteristic times\cite{Kaplan}
and the sum over the three  branches is essential to conserve energy.\cite{Goldie_SuST_2013}

\section{The effect of a pair-breaking signal}
\label{sec:Pair_breaking_signal}
The coupled kinetic equations described in Ref.~\onlinecite{Chang_and_Scalapino_ltp}
 were solved using
 Newton-Raphson iteration to find the non-equilibrium quasiparticle and phonon energy distributions
 $f(E)$ and $n(\Omega)$.
Details of the scheme are given in
Ref.~\onlinecite{Goldie_SuST_2013}.
The absorbed powers per unit volume from the signal $P_{s}$ and probe  $P_{p}$  are assumed  to be
spatially uniform.
We ignore changes in  $\Delta$ due to $P_{s}$ and  $P_{p}$.
In Ref.~\onlinecite{de_Visser_Goldie_2013} we found that changes in $\Delta$ were very small $\ll 0.001\Delta$
for typical experimental $P_{p}$.
%
%
%
%
%
%
The effect of 
$P_{s}$
is to introduce an additional drive term\cite{Eliashberg_1} into Eq.~[2] of Ref.~\onlinecite{Goldie_SuST_2013}
for the quasiparticle distribution function
%
%
$ \delta f(E)/\delta t \vert_{s} = I_{s}  $
where $I_{s}=B_{s}K_{s}$,
\begin{equation}
\begin{split}
K_{s}(E,\nu_s) = K_p(E,\nu_s)
			& +2\rho(E^{\prime} ,\Delta) \left[ 1- \frac {\Delta^2} {E E^{\prime} } \right]\\
 & \times \left[1 - f\left(E \right) -f\left( E^{\prime} \right) \right],
\label{Eq:Drive_1}
\end{split}
\end{equation}
 $E^{\prime} = h\nu_{s}- E $ and the prefactor $B_{s}$ is calculated with
$B_{s} = P_{s}/4 N(0) \int_\Delta^\infty E \rho(E) K_{s}(E,\nu_{s}) dE$,
 ensuring that  the  absorption of $P_s$ conserves energy.
A prefactor for the probe power absorption $B_p$ can be similarly defined.\cite{Goldie_SuST_2013}
$N(0)$ is the single-spin electronic density of states.
%
%
%
%
Some  solutions require calculation of differences between
 distributions. To ensure numerical accuracy we increased the precision requirements
 in the
code so that the errors in the power flow between the  quasiparticles and thin-film phonons and
then the heat bath phonons
(see Ref.~\onlinecite{Goldie_SuST_2013} for details)
were converged to better than $2\times10^{-6}$ 
and likewise the iterated solutions for  $B_s$ and $B_p$.
  We used a  quasiparticle density of states
$\rho(E,\Delta)= {\rm Re}  \left(   \left( E+i\gamma \right) / \left( ( E+i\gamma \ )^2-\Delta^2 \right)^{1/2}  \right)  $.
The factor $\gamma$ takes account of the broadening of the peak in $\rho$ near $E=\Delta$ due to lifetime effects or
film inhomogeneity.\cite{Jonas_review}
 The choice $\gamma= 1.125 \times 10^{-3}\Delta $
minimizes the difference between the
thermal quasiparticle number density $N_T$ calculated by summing over the discretized distributions (where we used a $1\,\,{\rm \mu eV}$ grid)
compared to numerical integration of the functions $N_T=4 N(0)\int_\Delta^\infty \rho(E,\Delta)f(E,T) dE$
where $f(E,T)=1/\left(1+\exp(E/k_bT)\right)$ is the Fermi-Dirac function and $k_b$ is Boltzmann's constant.
%
%
%
%
%
We  used parameters of a thin Al  film   as in Ref.~\onlinecite{Goldie_SuST_2013}:
 $\Delta =180\,\,{\rm \mu eV}$, $T_c=1.17\,\,{\rm K}$,
$N(0) =1.74\times 10^{4} \,{\rm \mu eV^{-1} \mu m^{-3}}$,
characteristic quasiparticle time
$\tau_0=438\,\,{\rm ns}$.\cite{Kaplan}
 $\tau_0^\phi=0.26\,\,{\rm ns}$   and  $T/T_c=0.1$.
 This ratio of $\tau_0/\tau_0^\phi $ means that our numerical solutions conserve energy,
they are {\it not} independent variables: Eq.~[11] of Ref.~{\onlinecite{Goldie_SuST_2013}} gives the overall
 parameter dependencies.
The value we use for $\tau_0$ has given a  good account of the temperature
dependence of the generation-recombination noise measured in clean, thin Al films.\cite{deVisser_prl, de_Visser_Goldie_2013}
in which the effect of
phonon trapping should be small (we estimate $\tau_l/\tau_0^\phi\sim 0.5$ in this case).
 A number of previous authors have used $\tau_0\sim 100\,\,{\rm ns}$,\cite{Chi_and_Clarke} although this
value seems inconsistent with the more recent measurements.
Wilson and Prober have also observed unexpectedly long lifetimes in $200\,\,{\rm nm}$  Al films
 (estimating $\tau_0$ to be even longer
than the value used here), and suggested  the observation
resulted from  an anomalously long $\tau_l$.\cite{Wilson_noise_2004}
Interestingly the longer $\tau_0$ seems to be associated with those measurements that have implemented
stringent experimental procedures to minimize the effect of stray light from higher temperature
stages in cryogenic systems: note that typical photon energies emitted by a $4\,\,{\rm K}$ source significantly exceed $2\Delta$ in Al.
%
%
Where used we assumed 
$h\nu_{p}=16\,\,{\rm \mu eV}$ ($\nu_{p}= 3.88\,\,{\rm GHz}$).
\section{Calculating $\eta_s$ }
Consider   $m$,  the average number  of driven quasistatic quasiparticles generated  by each
 absorbed photon.
Signal photons interact with rate $\Gamma_\Phi=P_s/h\nu_s$ and each photon creates
two primary quasiparticles. These rapidly relax in energy
generating  the driven quasistatic population with rate
$\Gamma_s=m\Gamma_{\Phi}$.
Assuming that all of the
excess quasiparticles have $E=\Delta$ then 
$\eta_{s} =m \Delta/h\nu_{s} = \Gamma_s \Delta/ P_s$.
We use a modified
set of Rothwarf-Taylor rate equations\cite{Rothwarf_Taylor} to find $\Gamma_s$.
With $\Gamma_p$ the generation rate of quasistatic quasiparticles
due to the probe, $N$  the number density of quasiparticles and $N_{2\Delta}$ the number density of
 $2\Delta$-phonons
\begin{equation}
\frac{d N}{d t}= \Gamma_{s}+\Gamma_{p} - RN^2+2\beta N_{2\Delta},
\label{Eq:RT_1}
\end{equation}
\begin{equation}
\frac{d N_{2\Delta} }{d t}= \frac{RN^2}{2} -\beta N_{2\Delta}-\frac{ N_{2\Delta} -N_{2\Delta}^{T}}{\tau_l}.
\label{Eq:RT_2}
\end{equation}
Here $R$ and $\beta$ are the recombination and pair-breaking coefficients respectively
and $ N_{2\Delta}^{T}$ is the
thermal density of $2\Delta$-phonons.
We assume that $\Gamma_s$ and $\Gamma_p$  are independent.
With $\Gamma_{s}=0$, Eqs.~\ref{Eq:RT_1} and \ref{Eq:RT_2}
can be solved by first also setting   $\Gamma_{p}=0 $ so that
in steady-state, $ d N / d t = d N_{2\Delta} /d t =0 $, giving
$R  N_{T} ^2/2=\beta N_{2\Delta}^T$.  This leads to
 $\Gamma_{p}=R\left(N_{p}^2 - N_{T}^2\right)/ \left( \beta\tau_l+1 \right)$,
where $N_{p}$ is the total number density of quasiparticles with the probe.
For the additional  signal  $\Gamma_{s}$ we find
\begin{equation}
\Gamma_{s}=R\left(N^2 - N_{p}^2\right)  \frac{1}{\beta\tau_l+1}.
\label{Eq:RT_33}
\end{equation}
%
%
and with $\beta=1/\tau_{pb}$
\begin{equation}
\eta_{s} = \frac { R \Delta  \left(N^2 - N_{p}^2\right)}{P_{s}}   \frac{1}{\tau_l/\tau_{pb}+1}.
\label{Eq:The_first_answer}
\end{equation}
%
$N$ and $N_{p}$ were calculated by numerically integrating solutions of the coupled kinetic equations.
We used Eq.~(A9)  of Chang and Scalapino\cite{Chang_and_Scalapino_prb} to
define a recombination rate  $R_{\rm CS}$. We find that setting $R\equiv 2R_{\rm CS}$
 ensures that the population-averaged recombination time $\langle \tau_r\rangle_{qp}=1/RN$ in thermal equilibrium
($N=N_T$)
is the same calculated using either Refs.~\onlinecite{Chang_and_Scalapino_prb} or \onlinecite{Kaplan}.
We calculate  $\langle \tau_{pb}\rangle_\phi$ for $f(E)$ and $n(\Omega)$  using Eq.~(A10) of  Ref.~\onlinecite{Chang_and_Scalapino_prb}.
%
Writing Eq.~\ref{Eq:RT_33}
in terms of the excess number densities $N_s^{ex}$ due to the signal and $N_p^{ex}$ due to the probe alone so that
 $N=N_s^{ex}+N_p^{ex}+ N_T$ and $N_p=N_p^{ex}+N_T$ the effective recombination time  $\tau_r^{\rm {eff}}=N_{s}^{ex}/\Gamma_s$
can be calculated for any combination of the  magnitudes
of $N_s^{ex}$, $N_p^{ex}$ and $N_T$. In the calculations reported we consider signal and probe powers relevant to
ultra-sensitive KIDs for astronomical applications\cite{de_Visser_Goldie_2013}
so that
 $N,\,\,N_p \gg N_T$ for all cases of $P_p$, $P_s$  studied.
\begin{figure}[!t]
 \begin{center}
   \begin{tabular}{c}
    \psfrag{Xaxis} [] [] { $E /\Delta  $ }
    \psfrag{Yaxis} [] [] { $ f(E)  $  }
    \psfrag{Yaxis2} [] [] [0.85] { $ K_{s} \rho(E,\Delta)\,{\rm (Hz)}  $   }
     \psfrag{aa} [r] [r] [0.95] { $h\nu_{s} -\Delta $ }
   \psfrag{bb} [r] [r] [0.95] { $h\nu_{s} + \Delta $ }
    \psfrag{cc} [r] [r] [0.9] { $h\nu_{s} - \Delta $ }
   \includegraphics[height=8.5cm, angle=-90]{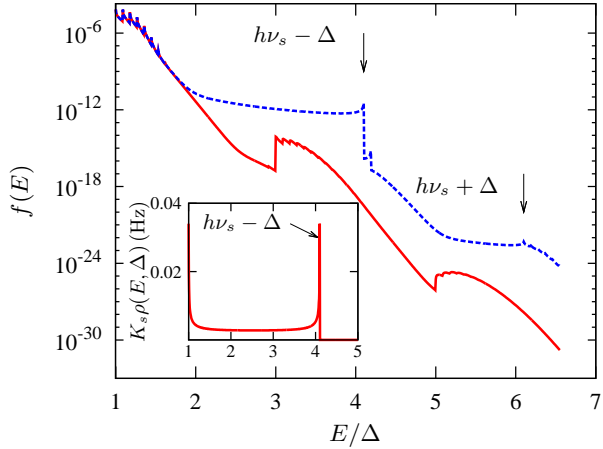}
   \end{tabular}
  \end{center}
  \caption[Fig1]
   { \label{fig:f_with_signal}
 (Color online) Semi-log plot showing the effect of   $P_{p}=20\,\,{\rm aW/\mu m^3}$ on the quasiparticle distribution with
$T/T_c=0.1$:
  (full red line)  probe power only and  (dashed blue line) with additional signal $P_{s}/P_{p}=0.01$.
The  signal photon energy $h\nu_{s}=5.1\Delta$ and $\tau_{l}/\tau_0^\phi = 1$. The inset
 shows the contribution to the number drive $K_{s} \rho(E,\Delta)$
for the signal normalized so that each absorbed photon produces two primary quasiparticles per second.
 }
   \end{figure}
%
%
%
%
\section{Results}
\label{sec:Results}
\begin{figure}[!t]
 \begin{center}
  \begin{tabular}{c}
   \psfrag{Xaxis} [] [] { $\Omega/ \Delta  $ }
	\psfrag{Yaxis} [] [] { ${ \delta P(\Omega)_{\phi-b}/d\Omega }\,\, ( \rm{ MHz/  \mu m^3} )  $ }
   \psfrag{A} [r] [r] [0.85] { $P_{s}=0.1P_{p} $ }
   \psfrag{B} [r] [r] [0.85] { $P_{s}=0.01P_{p}  $ }
   \psfrag{ab} [l] [l] [0.95] { $\Omega= h\nu_{s} - 2 \Delta $ }
   \psfrag{bb} [r] [r] [0.95] { $\Omega= h\nu_{s} $ }
   \includegraphics[height=8.5cm, angle=-90]{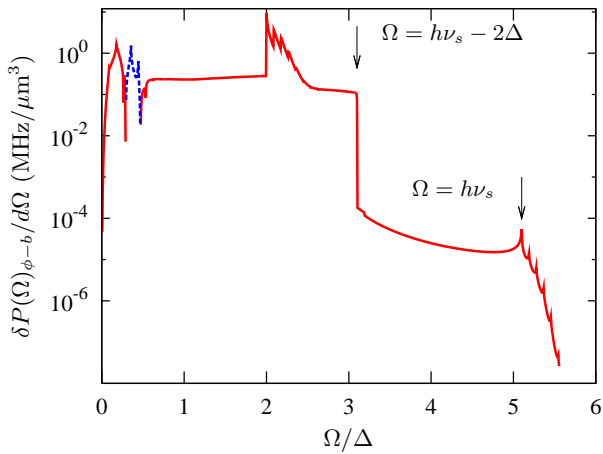}
   \end{tabular}
  \end{center}
   \caption[Fig3]
   { \label{fig:dP_Phonon_sig}
 (Color online) The change in the phonon power flow to the bath $\delta P(\Omega)_{\phi-b}$ for
    $P_{s}=0.01P_{p} $ and $P_{p}=20\,\,{\rm aW/\mu m^3}$.
   The  signal photon energy $h\nu_{s}=5.1\Delta$ and $\tau_{l}/\tau_0^\phi = 1$.
The (blue) dashed section indicates the phonon energies for which the change  is negative.
 }
   \end{figure}
%
 Fig.~\ref{fig:f_with_signal} shows  $f(E)$ for
$P_{p}=20\,\,{\rm aW/\mu m^3}$, as the solid curve
and the additional effect of  $P_{s}/P_{p}=0.01  $ (dashed blue curve) with $h\nu_{s}=5.1\Delta$.
The inset  
 shows the contribution to the number drive $K_{s} \rho(E,\Delta)$
for the signal normalized so that each absorbed photon produces two quasiparticles per unit time.
The double peak arises because $K_{s} \rho(E,\Delta)$ involves the {\it product}
of final state densities $\rho(E,\Delta)\rho(E^\prime,\Delta)$ which
 is symmetric
with respect to the final state energies.
The main figure shows that $f(E)$ for the probe alone   has  multiply
peaked structure at  $E \sim \Delta $ due to absorption of the probe photons by the large density of quasiparticles near $\Delta$.
At energy $E = 3 \Delta $ there is a step in $f(E)$  corresponding to reabsorption of $2\Delta$-phonons by the driven
quasiparticles that also exhibits peaks associated with multiple photon absorption from the probe. A smaller feature
at $E=3\Delta - h\nu_p$ is  visible that arises from stimulated emission.

The dashed curve showing $f(E)$ with $P_s$ has similar structure at low energies but now  shows a
step  at $E=h\nu_{s}-\Delta$.
The curvature $f(E)$ below this primary peak arises from the energy dependence of the
quasiparticle scattering and recombination rates. The peak also has a smaller ``satellite" at
$E=h\nu_{s}-\Delta + h \nu_p $ as multiple photon processes  involving signal
  and probe occur.
A further similar feature is evident at $E=h\nu_{s}+\Delta$.
\begin{figure}[!t]
 \begin{center}
   \begin{tabular}{c}
   \psfrag{Xaxis} [] [] { $ h\nu_s/\Delta $ }
   \psfrag{Yaxis} [] [] { $ \eta_{s} $  }
   \psfrag{A} [r] [r] [0.85]  { $ \tau_l/\tau_{0}^\phi=0.5 $  }
   \psfrag{B} [r] [r] [0.85]  { $ 1.0 $  }
   \psfrag{C} [r] [r] [0.85]  { $ 2.0 $  }
   \psfrag{D} [r] [r] [0.85]  { $ 4.0 $  }
   \psfrag{E} [r] [r] [0.85]  { $ 8.0 $  }
   \includegraphics[height=8.5cm, angle=-90]{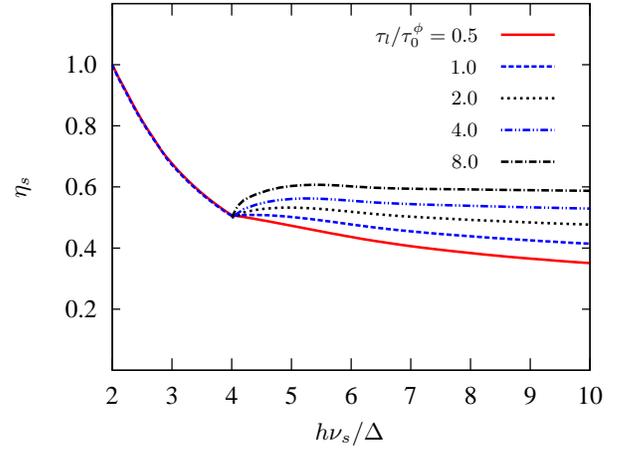}
   \end{tabular}
  \end{center}
   \caption[Fig1]
   { \label{fig:Eta_sig_tauloss}
 (Color online) Number generation efficiency $\eta_{s}$ as a function of $h\nu_s/\Delta$ for 5 values of  $\tau_{l}/\tau_{0}^\phi $.   }
   \end{figure}
%
%
\begin{figure}[!t]
 \begin{center}
   \begin{tabular}{c}
   \psfrag{Xaxis} [] [] { $ P_s\,\left({\rm aW /\mu m^3}\right)  $ }
   \psfrag{Yaxis} [] [] { $ \eta_{s} $  }
   \psfrag{label1} [l] [l] [0.85]  { $ P_p=20\,{\rm aW/\mu m^3}{\rm ,\,\,} h\nu_s=5\Delta $  }
   \psfrag{label2} [l] [l] [0.85] { $ P_p=0{\rm ,\,\,} h\nu_s=5\Delta$  }
   \psfrag{label4} [l] [l] [0.85]  { $ P_p=20\,{\rm aW/\mu m^3}{\rm ,\,\,} h\nu_s=3\Delta $  }
   \psfrag{label3} [l] [l] [0.85] { $ P_p=0{\rm ,\,\,} h\nu_s=3\Delta$  }
   \includegraphics[height=8.5cm, angle=-90]{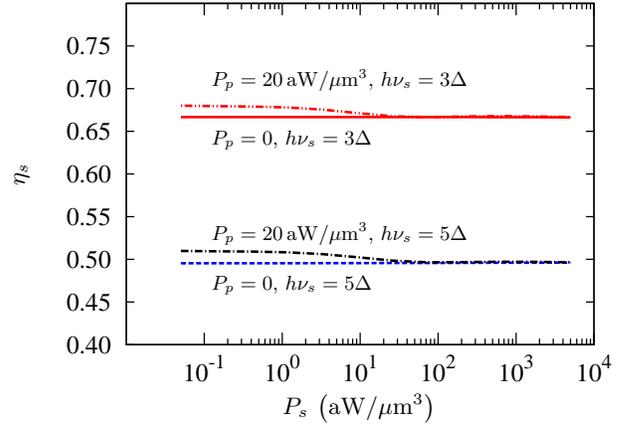}
   \end{tabular}
  \end{center}
   \caption[Fig1]
   { \label{fig:Eta_model}
 (Color online) Number generation efficiency $\eta_{s}$ for $h\nu_s=3 $ and $5\Delta$
 with  $P_{s}=0$ and $20 \,\,{\rm aW/\mu m^3}$  with $\tau_{l}/\tau_0^\phi = 1$. }
   \end{figure}

Fig.~\ref{fig:dP_Phonon_sig} shows
the change in contributions to the power flow to the heat bath
$\delta P(\Omega)_{\phi-b}= P(\Omega)^{s}_{\phi-b}- P(\Omega)^{p}_{\phi-b}$, where
$P(\Omega)^{s}_{\phi-b}$ is the contribution to the phonon-bath power flow with signal and probe, and $P(\Omega)^{p}_{\phi-b}$
that for the probe alone.
At low phonon
energies $\Omega<0.3  \Delta $,  $\delta P(\Omega)_{\phi-b}$   is  increased  due to  pair-breaking. At energies
$0.3< \Omega < 0.5 \Delta $ the net flow is negative. The first effect arises as the signal itself has a sharply peaked structure
near the gap. The reduction 
 arises from the blocking of final states for the  scattering of higher energy
probe-generated quasiparticles  towards the gap.
At higher phonon energies there is a significant change in $\delta P(\Omega)_{\phi-b}$   due to
 phonons $\Omega\ge 2\Delta $. 
The spectrum also shows a broad low background contribution at all phonon energies
$\Omega\le \left(h\nu_{s}-2\Delta \right)$
generated as the primary spectrum scatters to energies $E\sim \Delta$
and at higher $\Omega$ from the highest energy quasiparticles shown in Fig.~\ref{fig:f_with_signal}.
%
%

%

Fig.~\ref{fig:Eta_sig_tauloss}
shows calculations of $\eta_{s} $ as a function of $h\nu_s$ for 5 values of
$ \tau_l/\tau_{0}^\phi$. The calculation used $P_s=0.2\,{\rm aW/\mu m^3}$ and $P_p=20\,{\rm aW/\mu m^3}$.
For $2\Delta \le h\nu_s \le 4\Delta$, $\eta_s$ reduces monotonically
and is independent of the phonon loss time.
In this regime the high energy primary quasiparticle peak is created at $\Delta\le E\le 3\Delta$ and  phonons emitted in scattering
are unable to break pairs.
At higher signal energies $4\Delta \le h\nu_s \le 6\Delta$ the efficiency tends to increase again and the increase depends on
$\tau_{l}/\tau_{0}^\phi $.
Pair-breaking enhances $\eta_s$ and the enhancement depends on the probability of pair-breaking compared to other phonon losses.
At higher energies multiple pair-breaking is necessary to create the low energy steady state distribution, but multiple
phonon loss also occurs. The overall effect is a reduction in $\eta_s$ as $h\nu_s$ increases for finite $\tau_{l}/\tau_{0}^\phi $.
We note that $\eta_s \to 1$ as $h\nu_s\to 2\Delta$ for all $ \tau_l/\tau_{0}^\phi$ so that
Eqs.~(\ref{Eq:RT_1}) to (\ref{Eq:The_first_answer})
and our definition of $R$ self-consistently conserve energy.

Fig.~\ref{fig:Eta_model}
shows  $\eta_{s}$ for
$h\nu_s=3{\rm \, and\,\,}5\Delta$
as a function of $P_{s}$ for two values of the probe power.
For $P_p=0$ the generation efficiency is constant over 5 orders of magnitude of absorbed signal powers.
For $P_p=20\,\,{\rm aW/\mu m^3}$ and for low signal power, $\eta_{s}$ is slightly enhanced.
We discuss this in the Sec.~\ref{Sec:Discussion}.
%
%
\begin{figure}[!t]
 \begin{center}
   \begin{tabular}{c}
   \psfrag{Xaxis} [] [] { $ P_s\,\left({\rm aW /\mu m^3}\right)  $ }
   \psfrag{Yaxis} [] [] { $ \tau_{r}\,\,({\rm \mu s})  $  }
   \psfrag{a} [r] [r] [0.8] { $ P_p=0\,\, h\nu_s=3\Delta   $  }
   \psfrag{b} [r] [r] [0.8] { $ h\nu_s =5\Delta   $  }
   \psfrag{c} [r] [r] [0.8]  { $ P_p=20\,{\rm aW/\mu m^3} \,\, h\nu_s=3\Delta   $  }
   \psfrag{d} [r] [r] [0.8]  { $ h\nu_s=5\Delta   $  }
   \psfrag{Pp} [l] [l] [0.8]  { $ P_p=20\,{\rm aW/\mu m^3}    $  }
   \psfrag{Psonly} [l] [l] [0.8]  { $ P_s$ only }
   \psfrag{Probe} [l] [l] [0.8]  { With probe }
   \includegraphics[height=8.5cm, angle=-90]{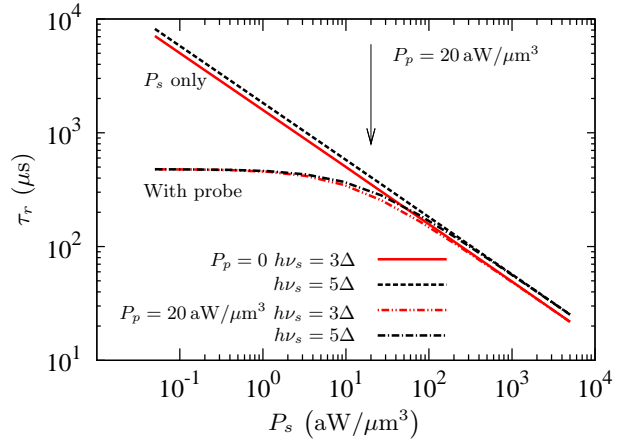}
   \end{tabular}
  \end{center}
   \caption[Fig1]
   { \label{fig:taur_model}
 (Color online) Distribution-averaged recombination times associated with the calculations of $\eta_s$ shown in Fig.~\ref{fig:Eta_model}:
(red) solid and double-dot dashed lines $h\nu_s=3\Delta$, (black) dashed and dot-dashed lines $h\nu_s=5\Delta$ with the
values of $P_p$ as indicated.    }
   \end{figure}
%
%
\begin{figure}[!t]
 \begin{center}
   \begin{tabular}{c}
   \psfrag{Xaxis} [] [] { $ P_s\,\left({\rm aW /\mu m^3}\right)  $ }
   \psfrag{Yaxis} [] [] { $ \tau_{pb}/\tau_0^\phi   $  }
   \psfrag{a} [l] [l] [0.85]  { $ P_p=0 {\rm ,\,\,} h\nu_s=3\Delta$  }
   \psfrag{b} [l] [l] [0.85] { $ P_p=20\,{\rm aW/\mu m^3}   $  }
   \psfrag{c} [l] [l] [0.85]  { $ P_p=0{\rm ,\,\,} h\nu_s=5\Delta$  }
   \psfrag{d} [l] [l] [0.85]  { $5\Delta$  }
   \psfrag{e} [l] [l] [0.85]  { $3\Delta$  }
   \includegraphics[height=8.5cm, angle=-90]{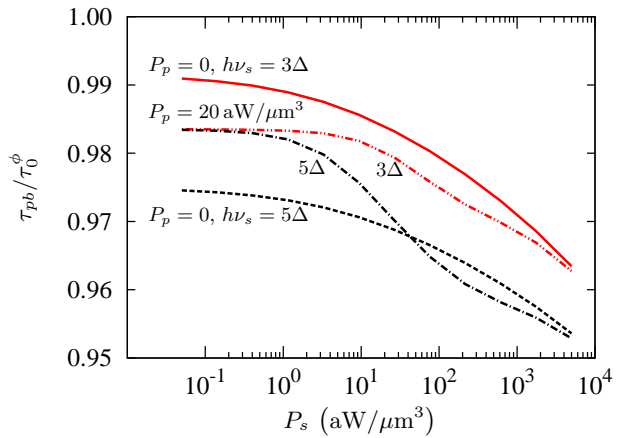}
   \end{tabular}
  \end{center}
   \caption[Fig1]
   { \label{fig:tpb_model}
 (Color online)  Distribution-averaged pair-breaking times used in the calculations of $\eta_{s} $ shown in Fig.~\ref{fig:Eta_model}:
(red) solid and double-dot dashed lines $h\nu_s=3\Delta$, (black) dashed and dot-dashed lines $h\nu_s=5\Delta$ with the
 values of $P_p$  indicated.    }
   \end{figure}
%
Fig.~\ref{fig:taur_model} shows  the distribution-averaged
values of $\tau_r$ associated with the the calculations shown in Fig.~\ref{fig:Eta_model}
(note that Eq.~\ref{Eq:The_first_answer} does not involve $\tau_r$ explicitly)
while
Fig.~\ref{fig:tpb_model}  shows   $\tau_{pb}$
that is directly
 used in these calculations. 
 Considering Fig.~\ref{fig:taur_model} for the small signal regime $P_s\ll P_p$,
 $N_p$ determines $\tau_r$.
 For the large signal regime $P_s\gg P_p$,
$\tau_r$ is independent of $P_p$ because $N_s^{ex}$ determines $\tau_r$.
 We  see that  for fixed $h\nu_s$,
 $\tau_r$  changes by nearly three orders of magnitude whilst $\eta_s$
 shown in Fig.~\ref{fig:Eta_model}
is  constant for all $P_s$.
 (Very close inspection of the results for $P_p=0$
 would show that $\eta_s$ varies for the range of calculated $P_s$
  by about $\pm 0.03\%$ which we consider acceptable given the numerical precision used and the discretization of the distributions.)
Also $\tau_r$ for the two signal photon energies differ, but the generation rate of excess quasiparticles
 depends on  $h\nu_s$, as does the fraction of power  lost before the quasistatic driven distributions are created.
We would note that Eq.~\ref{Eq:The_first_answer}
correctly takes into account  these underlying  changes in $\tau_r$ due to signal and probe.

Fig.~\ref{fig:tpb_model} shows  that the distribution-averaged $\tau_{pb}\sim \tau_0^{\phi}$, which would often be assumed, but moreover
$\tau_{pb} < \tau_0^{\phi}$. For $T/T_c=0.1$, $\tau_{pb}(\Omega)\le\tau_{pb}(2\Delta)$.
 $\tau_{pb}(\Omega)$ scales approximately as $ 1/\Omega$
in thermal equilibrium,\cite{Kaplan} hence the slight reduction
when $\tau_{pb}$ is calculated for the non-equlibrium distributions.
We find that the
variation of $\tau_{pb}$ as a function of the drive (both probe and signal) arises from the detailed
spectra of the $2\Delta$-phonons for each case and  is (to first-order) independent of the
 quasiparticle spectrum. It is possible to define an effective phonon temperature $T_{2\Delta}^{\rm {eff}}$ that accounts for  the total number
of $2\Delta$-phonons. This approach accounts for the calculated $\tau_{pb}$ to within $~1\%$,
but not for the detailed behavior as a function of $P_s$.
In the presence of the probe and signal the probe determines $\tau_{pb}$ if $P_p\gg P_s$
 and in this case we find $T_{2\Delta}^{\rm {eff}}$  more closely accounts for
the calculated $\tau_{pb}$. We would emphasize here that
the  full calculation of $\tau_{pb}$ is necessary to find that $\eta_s$ is independent
of power $P_s$.
\section{Discussion and Conclusions}
\label{Sec:Discussion}
We have presented  calculations of the quasiparticle generation efficiency $\eta_s$
for a pair-breaking signal in thin Al films at $T/T_c=0.1$ with photon
energies in the range $2\Delta\le h\nu_s\le 10\Delta$, $90\le \nu_s \le 450\,\,{\rm GHz}$.
We have also investigated the effect of including  a probe
 with power and frequency typical of those used in
 low-noise KID readout.
 The calculated detailed spectra  show the effects of multiple interactions of the probe and the signal
 in the driven $f(E)$
 with structure for example at $E=h(\nu_{s}+\nu_{p})$.
Our results demonstrate the importance of phonon loss on the  quasiparticle creation efficiency.
For thick films, $\tau_{l}/\tau_0^\phi = 8$, our calculations are in general agreement with earlier work for much higher signal
 energies, in calculations that ignore $2\Delta$-phonon loss, showing $\eta_s~\simeq 0.59$.
For resonators, thinner films would tend to be used since these maximize the kinetic inductance fraction of the response,\cite{Jonas_review}
but these have reduced creation efficiencies by as much as $40\%$ for the thinnest films considered here.
Our calculations establish limits on the detection sensitivity of thin-film superconductors.
The limiting Noise Equivalent Power of a thin-film detector is determined by generation-recombination noise\cite{Wilson_noise_2004,deVisser_jltp,Goldie_SuST_2013}
 and is
 given by $NEP=2\Delta\sqrt{N V/\tau_r^{\rm {eff}}}/\eta$ where $V$ is the volume of the film and $\eta$ the overall detection efficiency.
 $\eta$ is the product of all detection efficiencies (including coupling efficiency)  but $\eta_s$ shown in Fig.~\ref{fig:Eta_sig_tauloss}
determines the limiting efficiency in the thin-film case. Fig.~[8] of
Ref.~\onlinecite{Goldie_SuST_2013}
shows calculations of the limiting coupled ${\emph {NEP}} $ for $\eta=0.59$ and $1$
as a function of absorbed probe power. The present work shows that the best-possible
coupled ${\emph {NEPs}} $  are even higher  than the case $\eta=0.59$ shown there
for much of the mm- and sub-mm spectrum in thin superconducting films.
In deriving Eq.~\ref{Eq:The_first_answer} we assumed that all quasiparticles have energy $\Delta$. It is possible to take account of the
energy distribution of the excess quasiparticles in the derivation and this  would increase our calculated $\eta_s$
by about $4\%$, but for consistency with earlier work we have assumed $E=\Delta$ for all of the excess.

We identify a coupling between the signal and  probe that enhances $\eta_s$ by about $2\%$. 
This maybe the effect described by Gulian and van~Vechten,\cite{Gulian_van_Vechten}
who suggested that for low $P_s$ multiple probe photon absorption by the higher energy primary peak of Fig.~\ref{fig:f_with_signal}~(inset)
occurs and some fraction of these quasiparticles are driven to energies $E \ge 3 \Delta$.
By contrast
Fig.~\ref{fig:f_with_signal} suggests that $2\Delta$-phonon
reabsorption occurs 
to enhance $\eta_s$.
As the signal power increases there is a slight {\it reduction} in $\eta_{s}$ because the relevant quantity is the fraction
of quasiparticles in the photon peak driven above the pair-breaking threshold. The fraction reduces because the
probe power is fixed and the probe generates (most of) the excess $2\Delta$-phonons.
In future work we intend to  extend the work to consider other low temperature superconductors, to investigate the detection linearity of a resonator with the
driven distributions and also to consider the  probe power levels that optimize detector $NEPs$.




\bibliographystyle{hunsrt}
\bibliography{TESreferences5}   

\end{document}